\documentclass[
reprint,
superscriptaddress,
amsmath,amssymb,
aps,
showpacs,
prx,
]{revtex4-2}

\usepackage{CJK}
\usepackage{color}
\usepackage{subcaption}
\usepackage{graphicx}
\usepackage{caption}
\usepackage{dcolumn}
\usepackage{bm}

\usepackage[mathlines]{lineno}
\usepackage{dsfont}
\usepackage{xcolor}
\usepackage{numprint}
\usepackage{siunitx}
\npthousandsep{,}
\usepackage{mathbbol}
\usepackage{amssymb} 
\usepackage{subcaption}
\usepackage{graphicx}
\usepackage[T1]{fontenc}

\DeclareSymbolFontAlphabet{\amsmathbb}{AMSb}

\newcommand{\ket}[1]{\mathop{| #1 \rangle}\nolimits}

\begin{document}
\begin{CJK*}{UTF8}{gbsn} 
\preprint{APS/123-QED}
\title{Broad Instantaneous Bandwidth Microwave Spectrum Analyzer \\with a Microfabricated Atomic Vapor Cell}

\author{Yongqi Shi (石永麒)}
\email{yongqi.shi@unibas.ch}
\affiliation{Department of Physics, University of Basel, Klingelbergstrasse 82, 4056 Basel, Switzerland}

\author{Thomas Ruster}
\author{Melvyn Ho}
\affiliation{Rohde \& Schwarz GmbH \& Co. KG, M{\"u}hldorfstra{\ss}e 15, 81671 Munich, Germany}

\author{Sylvain Karlen}
\author{Jacques Haesler}
\affiliation{Centre Suisse d'\'Electronique et de Microtechnique (CSEM) SA, Rue Jaquet-Droz 1, 2002 Neuch\^{a}tel, Switzerland}%

\author{Philipp Treutlein}
\email{philipp.treutlein@unibas.ch}
\affiliation{Department of Physics, University of Basel, Klingelbergstrasse 82, 4056 Basel, Switzerland}
\date{9 August 2024}

\begin{abstract}
We report on broad instantaneous bandwidth microwave spectrum analysis with hot $^{87}$Rb atoms in a microfabricated vapor cell in a large magnetic field gradient. The sensor is a MEMS atomic vapor cell filled with isotopically pure $^{87}$Rb and $\mathrm{N}_2$ buffer gas to localize the motion of the atoms.
The microwave signals of interest are coupled through a coplanar waveguide to the cell, inducing spin flip transitions between optically pumped ground states of the atoms. 
A static magnetic field with large gradient maps the \textit{frequency spectrum} of the input microwave signals to a position-dependent \textit{spin-flip pattern} on absorption images of the cell recorded with a laser beam onto a camera. In our proof-of-principle experiment, we demonstrate a microwave spectrum analyzer that has $\approx 1$~GHz instantaneous bandwidth centered around 13~GHz, 3~MHz frequency resolution, $2$~kHz refresh rate, and a -23~dBm single-tone microwave power detection limit in 1~s measurement time. A theoretical model is constructed to simulate the image signals by considering the processes of optical pumping, microwave interaction, diffusion of $^{87}$Rb atoms, and laser absorption. We expect to reach more than 25~GHz instantaneous bandwidth in an optimized setup, limited by the applied magnetic field gradient. Our demonstration offers a practical alternative to conventional microwave spectrum analyzers based on electronic heterodyne detection.
\end{abstract}

\maketitle
\end{CJK*}

\section{\label{sec1:intro}Introduction}

More than a century after the wireless transmission experiments of Heinrich Hertz, the continuous development of radio transmission techniques has enabled a myriad of applications in the microwave (MW) regime ($0.3$ - $300$~GHz), including global navigation satellite system (GNSS) positioning, mobile data transmission, and ultra-wideband (UWB) sensing. Across many use-cases, assessing radio signals without prior assumptions on the signal spectrum and in a real-time manner requires a broad frequency capability \cite{da_Silva_RealTime_2013}. However, achieving broadband operation still proves difficult, with the latest time-overlapped fast Fourier transform (FFT) techniques providing usually several hundreds of MHz and only recently up to several GHz of instantaneous span \cite{rohde__schwarz_implementation_2015,keysight_technologies_Spectrum_Analysis_2020,tektronix_fundamentals_nodate}, limited by the present-day analog-digital converters (ADCs) \cite{walden_analog--digital_2008} and the resultant FFT processing burden. In order to bypass the conventional swept-tuned heterodyne configuration and FFT procedures, multiple analog methods have been proposed and demonstrated based on special antenna structures \cite{gupta_microwave_2009}, various photonic techniques \cite{zou_photonics_2016,pelusi_photonic-chip-based_2009,konatham_real-time_2020}, or the spectral hole burning of ion-doped crystals in cryogenic conditions \cite{menager_demonstration_2001,merkel_extreme_2016,louchet-chauvet_telecom_2020}. 

Recently, quantum spectrum analyzers are being investigated for MW spectrum analysis, which harness quantum systems such as atoms \cite{meyer_waveguide-coupled_2021,compton_wideband_2021,bounds2024quantum} or color centers in solids \cite{chipaux_wide_2015,shao_wide-field_2016,rosskopf2017quantum,magaletti_quantum_2022,magaletti2021ensemble,carmiggelt2023broadband,liumicrowave} as microscopic antennae that can be controlled by light on the quantum level and that convert the MW signal of interest into an optical response. Such quantum approaches can potentially cover a wide carrier frequency range from MHz to THz, offer large instantaneous bandwidth, high sensitivity, and traceability to SI units. 

Thanks to their ease of use and their potential for low-cost, high-volume production \cite{Kitching_2018}, atomic vapor cells containing alkali species are one of the main candidates for commercial quantum sensors \cite{kitching_atomic_2011}, which have already been employed in atomic clocks  \cite{knappe_microfabricated_2004}, magnetometers \cite{budker_optical_2007}, gyroscopes \cite{kornack_nuclear_2005} and gas sensors \cite{schmidt_optogalvanic_2020}.  In recent years, atomic vapor cells are also being investigated for detection and imaging of high-frequency electromagnetic fields, using either ground-state atoms to measure the magnetic field component \cite{bohi_simple_2012,horsley_widefield_2015,horsley_frequency-tunable_2016,sun_rabi_2018,gerginov_atomic_2019} or Rydberg atoms to measure the electric field component \cite{sedlacek_microwave_2012,fan_subwavelength_2014,jing_atomic_2020}, from MW up to THz frequencies  \cite{downes_full-field_2020,chen_terahertz_2022}. The quantum nature of atoms enables measurements of the amplitude or frequency of the electromagnetic field based on well-known atomic properties and fundamental constants, providing intrinsic calibration while at the same time suffering less from crosstalk and electromagnetic interference than conventional antenna systems. While most work has focused on detecting electromagnetic field strength, early experiments have observed frequency information of a sub-MHz-band radio-frequency wave with alkali vapor \cite{alzetta_experimental_1976,tam_optically_1977,tam_optical_1979,davydov_quantum_2001,fescenko_imaging_2014}. Recent work on a Rydberg atom MW receiver demonstrated a real-time bandwidth of 16~MHz \cite{borowka2024continuous}. Measurements of the frequency spectrum of a broadband MW signal have not yet been demonstrated with atomic vapor cells.

In this work we report on the experimental demonstration and theoretical study of a microwave frequency spectrum analyzer based on ground state alkali atoms in a micro-electro-mechanical systems (MEMS) atomic vapor cell fabricated at wafer-level. Our spectrum analyzer features a broad instantaneous bandwidth and is non-cryogenic, easy to align, local oscillator-free, frequency-continuous, and only minimally perturbs the incident MW signal. 
After a brief introduction of the operating principle and experimental setup, we show measurement results for a single-tone MW signal and compare it to a theoretical simulation of the atomic spin dynamics and interactions. Various characterization measurements are presented to show the performance of our atomic microwave spectrum analyzer. We demonstrate its capabilities by recording the frequency spectra of a set of frequency-modulated (FM) MW signals and a time-dependent spectrogram of a frequency-swept signal. We discuss the dominant factors influencing several key specifications and present parameters for an optimized setup.

\section{\label{sec2:Principle} Principle}

\begin{figure*}[ht]
\includegraphics[width=2\columnwidth]{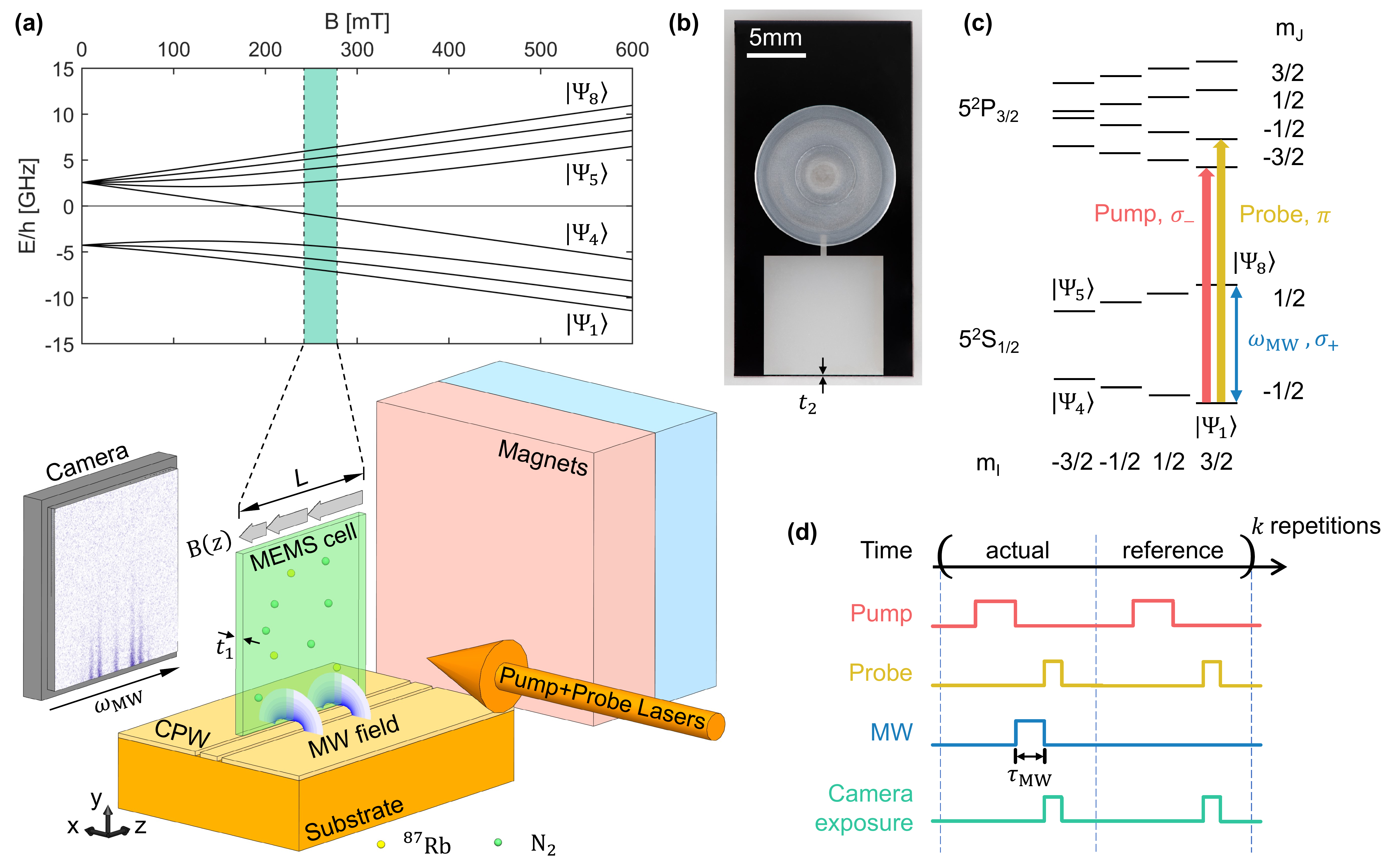}
\caption{\label{fig:principle}Setup and operation of the atomic MW spectrum analyzer. (a) Bottom: An illustration of the key components of the experimental setup. The interior of the MEMS cell containing the atomic vapor is depicted in green, with the long edge positioned directly above the signal trace of the CPW. Blue fan-shaped contours illustrate the $\sigma^+$ magnetic component of the MW near field in the $xy$-plane above the CPW. Permanent magnets are positioned so as to generate a large magnetic field with gradient along the $z$-axis. The pump and probe laser beams both propagate in the $x$ direction. The camera displays the atomic spin-flip signal recorded by absorption imaging, reflecting the frequency spectrum of the MW signal on the CPW. Top: Hyperfine energy levels of the $^{87}$Rb ground state as a function of the static magnetic field, with the region across the cell highlighted. (b) Photograph of a MEMS atomic vapor cell employed in the experiment, with a square sensing area of $10\,\mathrm{mm}$ side length and internal thickness $t_1 = 200\,\mu$m, filled with $^{87}$Rb vapor. The thin sidewall with $t_2 = 200\,\mu$m allows for a close approach of the CPW. (c) Hyperfine energy levels of the $5^2S_{1/2}$ ground and $5^2P_{3/2}$ excited states of $^{87}$Rb at a fixed position in the cell, with the relevant optical and microwave transitions indicated. (d) Experimental pulse sequence (see text).}
\end{figure*}

Our atomic spectrum analyzer exploits the ground-state spins of rubidium atoms as microscopic antennae, which resonantly interact with microwave radiation, and which can be initialized and read out with laser light. Their resonance frequency is made position-dependent with a static magnetic field gradient, so that the atoms effectively act as a large array of high-resolution frequency discriminators in parallel. The spatially dependent atomic response of this array to an applied microwave field is imaged with a laser onto a camera, revealing the frequency spectrum of the microwave signal.

To illustrate the working principle, consider an ensemble of effective spin-1/2 atoms in a long and thin vapor cell with buffer gas, placed close to a microwave coplanar waveguide (CPW) and subject to a static magnetic field $B(z) = B_0 + Gz$ with gradient $G$ along the $z$-direction, see Fig.~\ref{fig:principle}(a). The static field gradient results in a position-dependent spin-flip frequency $\omega(z) = \omega_0 + g \mu_B G z / \hbar$, where $\mu_B$ is the Bohr magneton, $g$ the Land{\'e} factor, $\hbar$ the reduced Planck constant, and $\omega_0$ contains contributions from the hyperfine splitting and the static offset field. When a microwave signal of angular frequency $\omega_\mathrm{MW}$ is applied to the CPW, its magnetic field component will induce spin-flip transitions in atoms around the position 
\begin{equation}
z_{\omega_\mathrm{MW}} = \frac{\hbar}{g \mu_B G} (\omega_\mathrm{MW} - \omega_0)
\end{equation}
where the atoms interact resonantly with the microwave. Microwave signals of different frequencies $\omega_\mathrm{MW}$ will thus be mapped to an atomic response at different positions $z_{\omega_\mathrm{MW}}$ in the cell. 

The atomic spectrum analyzer is operated in a pulsed mode, where the atomic spins are first initialized by optical pumping with laser light, which induces a population imbalance $\Delta p_0$ of the spin states coupled by the MW field. At a specific $z_{\omega_\mathrm{MW}}$, the resonant frequency component of the microwave signal on the CPW couples the spin states with Rabi frequency $\Omega = \mu_\mathrm{MW} B_{\omega_\mathrm{MW}}/\hbar$, where $B_{\omega_\mathrm{MW}}$ is the magnetic field amplitude of the microwave signal in the frequency interval $[\omega_\mathrm{MW},\omega_\mathrm{MW}+\gamma ]$, $\mu_\mathrm{MW}$ is the transition magnetic dipole moment, and $\gamma$ the transition linewidth. The coupling induces atomic spin flips at a rate $R$, which for weak fields ($\Omega \ll \gamma$) can be obtained from a rate equation model as 
\begin{equation}
R = \frac{\Omega^2}{\gamma} = \frac{(\mu_\mathrm{MW} B_{\omega_\mathrm{MW}})^2}{\gamma \hbar^2}.
\end{equation}
A short MW pulse of duration $\tau_\mathrm{MW}$ thus induces a population change $\Delta p_\mathrm{MW} = \Delta p_0 R \tau_\mathrm{MW}$ on the spin-flip transition, proportional to the microwave power in the frequency interval $[\omega_\mathrm{MW},\omega_\mathrm{MW}+\gamma ]$. After the interaction, an imaging laser pulse is applied that is resonant with one of the spin-states to image the induced change in atomic absorption onto a camera. For an unknown microwave signal on the CPW, this image reveals its frequency spectrum, with different frequency components being mapped to different locations on the image along the direction of the magnetic field gradient, see Figure~\ref{fig:principle}(a).

For a cell of length $L$, the instantaneous bandwidth of the spectrum analyzer is
\begin{equation}\label{eq:bandwidth}
\Delta \omega_\mathrm{BW} = g \mu_B G L/\hbar,
\end{equation}
scaling with $G$. The spectral resolution is given by the linewidth 
\begin{equation}
\gamma = \gamma_0 + g \mu_B G \Delta z_D/\hbar
\end{equation}
of the atomic spin flip transition, where $\gamma_0$ is the intrinsic linewidth due to collisional spin relaxation, and the second term accounts for broadening due to diffusion of atoms in the gradient by a distance $\Delta z_D$ during the microwave pulse.
For a strong gradient where the linewidth is diffusion limited, the spectrum analyzer can discriminate $n = \Delta \omega_\mathrm{BW}/\gamma \approx L/\Delta z_D$ different frequency components in the microwave signal simultaneously.
As a simple estimate, consider a gradient of $G=50~\mathrm{T}/\mathrm{m}$ across a cell of $L=2~\mathrm{cm}$, resulting in an instantaneous bandwidth of $\Delta \omega_\mathrm{BW}/(2\pi) = 28~\mathrm{GHz}$. At a buffer gas pressure of $100$~mbar, a diffusion distance of order $\Delta z_D=20~\mu\mathrm{m}$ during a $1~\mu$s long microwave pulse results in a spectral resolution of $\gamma/(2\pi) = 30~\mathrm{MHz}$ and $n = 10^3$. 
This demonstrates the potential of this technique for broad instantaneous bandwidth operation with high frequency resolution.

\section{\label{sec3:setup} Setup and operation}

Figure~\ref{fig:principle}(a) illustrates the experimental setup. The vapor cell, static gradient field, CPW and camera are arranged such that the MW field is homogeneous along the analysis direction $z$. The camera images the $yz$ plane so that the images show the microwave frequency spectrum along $z$ and the spatial dependence of the MW field amplitude along $y$, while the thickness of the cell along the propagation direction $x$ of the imaging beam is approximately matched to the extension of the MW field.

The atomic vapor cell is a miniaturized cell fabricated using wafer-scale MEMS microfabrication techniques \cite{Karlen2017}.
It consists of a patterned silicon layer of thickness $t_1 = 200\,\mu$m, defining the internal thickness of the cell, sandwiched between two glass plates and sealed by anodic bonding. Experiments have been carried out using three different MEMS cells with transverse dimensions of $10\,\mathrm{mm}\times10\,\mathrm{mm}$ (shown in Fig.~\ref{fig:principle}(b)), $6\,\mathrm{mm}\times6\,\mathrm{mm}$, and $2\,\mathrm{mm}\times2\,\mathrm{mm}$, respectively. 
All cells feature a thin sidewall of $t_2=200\,\mu$m thickness, allowing the vapor to closely approach the CPW structure and to couple to the near-field of the guided microwave. 
The cells are filled using $\mathrm{RbN}_3$ UV-decomposition \cite{Karlen2017} with isotopically selected $^{87}\mathrm{Rb}$ atomic vapor and $\mathrm{N}_2$ buffer gas to slow down the atomic motion, with buffer gas pressures of $100\,$mbar, $110\,$mbar, and $127\,$mbar, respectively, determined from the collisional broadening and shift of the $^{87}$Rb D2 line in absorption spectroscopy at $T_\mathrm{fill}=130^\circ\mathrm{C}$ and compared with calculations using the Elecsus software \cite{zentile_elecsus_2015,keaveney_elecsus_2018}. The diffusion coefficient of $^{87}$Rb in $\mathrm{N}_2$ buffer gas is estimated with $D=D_0\frac{P_0}{P_\mathrm{N_2}}(\frac{T_\mathrm{set}}{T_0})^{3/2}$, where $D_0 = 0.1819 \,\mathrm{cm}^2/\mathrm{s}$ \cite{pouliot2021accurate} is the diffusion coefficient at $P_0 = 1  \,\mathrm{atm}$ and $T_0 = 50^\circ\mathrm{C}$, and $P_\mathrm{N_2}$ is the $\mathrm{N}_2$ pressure scaled to $T_0$. The oven that heats the cell and stabilizes its temperature at $T_\mathrm{set}$ is not shown in Fig.~\ref{fig:principle}(a).

The cell and CPW are in close proximity to the surface of a large Samarium-Cobalt (SmCo) permanent magnet, which produces a large static magnetic field together with a gradient along the $z$-direction in the cell region, $\bm{\mathrm{B}}(\mathbf{r})\approx \bm{\mathrm{B}}(z)= \left(B_0 + Gz\right)\hat{\bm{\mathrm{e}}}_z$, where $B_0\approx 260\,\textrm{mT}$ and $G\approx 3.6\,\textrm{T/m}$. The ground-state energy level splittings of $^{87}$Rb as a function of magnetic field in the cell are highlighted in the upper plot of Fig.~\ref{fig:principle}(a). 

Different MW signals to be analyzed are generated by a signal generator (with power $P_\mathrm{MW}$) and gated with a switch. After a pre-amplifier, a directional coupler, and a circulator, the MW signal is coupled with net gain $\approx 32\,\mathrm{dB}$ into the CPW via a non-magnetic end-launch SMA connector. At the end of the CPW, it is coupled out of the chip in the same way and sent into a $50\,\Omega$ termination. Alternatively, since the atomic vapor cell does not significantly attenuate the MW signal, the termination can be replaced with a coaxial cable to use the signal for further measurements.

The laser beam for optical pumping of the vapor is produced by an external cavity diode laser (ECDL) and pulsed by an acousto-optical modulator (AOM). The probe beam is generated by a distributed feedback laser (DFB) and pulsed by a second AOM in double-pass configuration to increase the extinction ratio. The two beams are expanded to cover the entire cell 
and propagate along $x$ with the pump beam slightly tilted around $z$ to prevent it from being captured by the camera. A 4\textit{f} lens system images the cell plane onto a fast scientific complementary metal-oxide-semiconductor (sCMOS) camera. 
Because the buffer gas-induced collisional broadening of optical transitions ($\sim\!1.5\, \mathrm{GHz}$) exceeds the magnetic field gradient induced frequency span of the optical transitions across the cell ($\sim 0.6\,\mathrm{GHz}$ for $\sigma_-$ and $\sim 0.1\,\mathrm{GHz}$ for $\pi$), the pump and probe beams can address all atoms in the cell for our current parameters. For larger $G$ in future experiments, multiple optical pumping and probing beams in different regions of the cell can be applied.

The hyperfine energy levels of the $^{87}$Rb D2 line at $780$\,nm and the optical and microwave transitions involved in the atomic spectrum analyzer operation are shown in Fig.~\ref{fig:principle}(c), for atoms at a fixed position $z_{\omega_\mathrm{MW}}$ in the cell. The levels are labeled by the magnetic quantum numbers of the electron spin $m_J$ and nuclear spin $m_I$ along the magnetic field axis, which are  approximately good quantum numbers in the strong static field. For the field strengths in our experiment, the $\sigma_+$ component of the MW field drives the hyperfine transition between $\ket{\Psi_1}$ and $\ket{\Psi_8}$ while the other MW polarization components are off-resonant \cite{horsley_frequency-tunable_2016}.

The experimental pulse sequence is shown in Fig.~\ref{fig:principle}(d). It starts with a $10 \,\mu$s optical pumping pulse with $150 \,\textrm{mW}/\textrm{cm}^2$ intensity, which drives the $\sigma_-$ optical transition from $\ket{\Psi_1}$ followed by collisional decay due to the buffer gas. This builds up a steady-state population imbalance between $\ket{\Psi_1}$ and $\ket{\Psi_8}$, changing the optical depth $\textrm{OD}$ of the vapor for the probe light by $\Delta\textrm{OD}_\textrm{pump}$ compared to the initially unpolarized vapor. Subsequently, the MW signal to be analyzed is pulsed on for the duration $\tau_\mathrm{MW}$, during which it drives the $\sigma_+$ hyperfine transition between $\ket{\Psi_1}$ and $\ket{\Psi_8}$, generating a further change $\Delta\text{OD}_\textrm{MW}$ on top of $\Delta\text{OD}_\textrm{pump}$. For our parameters, $\Delta\text{OD}_\textrm{MW} < \Delta\text{OD}_\textrm{pump}$ across the cell. 
Finally, a linearly polarized probe pulse of duration 1\,$\mu$s and intensity $I_\mathrm{in}\approx 1.5\,\textrm{mW}/\textrm{cm}^2$ is sent to the cell, tuned to resonance with the $\pi$ transition from $\ket{\Psi_1}$. The probe pulse is attenuated by absorption in the atomic vapor as
\begin{equation}
I_\mathrm{out}^\mathrm{act}(z,y)=I_\mathrm{in}(z,y)e^{-(\mathrm{OD}-\Delta\mathrm{OD}_\mathrm{pump}+\Delta\mathrm{OD}_\mathrm{MW})}
\end{equation}
and recorded as the actual image by the sCMOS camera. Afterwards, the same sequence is applied again, but with the MW signal turned off, generating a reference image 
\begin{equation}
I_\mathrm{out}^\mathrm{ref}(z,y)=I_\mathrm{in}(z,y)e^{-(\mathrm{OD}-\Delta\mathrm{OD}_\mathrm{pump})}.
\end{equation}
From each actual and reference image, a common dark image is subtracted, measured separately with all lasers turned off, correcting for the offset counts of the camera and a small amount of ambient stray light. 
From the corrected actual and reference images, an image of 
\begin{equation}
\Delta\mathrm{OD}_\mathrm{MW}(z,y) = -\ln(I_\mathrm{out}^\mathrm{act}(z,y)/I_\mathrm{out}^\mathrm{ref}(z,y))
\end{equation}
is obtained, which for weak MW fields is proportional to the microwave-induced population change $\Delta p_\mathrm{MW}(z,y)$ and thus provides an image of the microwave spectrum.
To enhance the signal-to-noise ratio (SNR), the actual-reference sequence cycle is typically repeated $k$ times. 

A typical image of $\Delta\mathrm{OD}_\mathrm{MW}$ is displayed on the camera in Fig. \ref{fig:principle}(a), showing a frequency modulated MW signal, where the fringe pattern on the image reveals the spectrum (carrier and multiple sidebands) of the MW signal. 

\section{\label{sec4:results} Results and analysis}

\begin{figure*}[htp!]
\includegraphics[width=\textwidth]{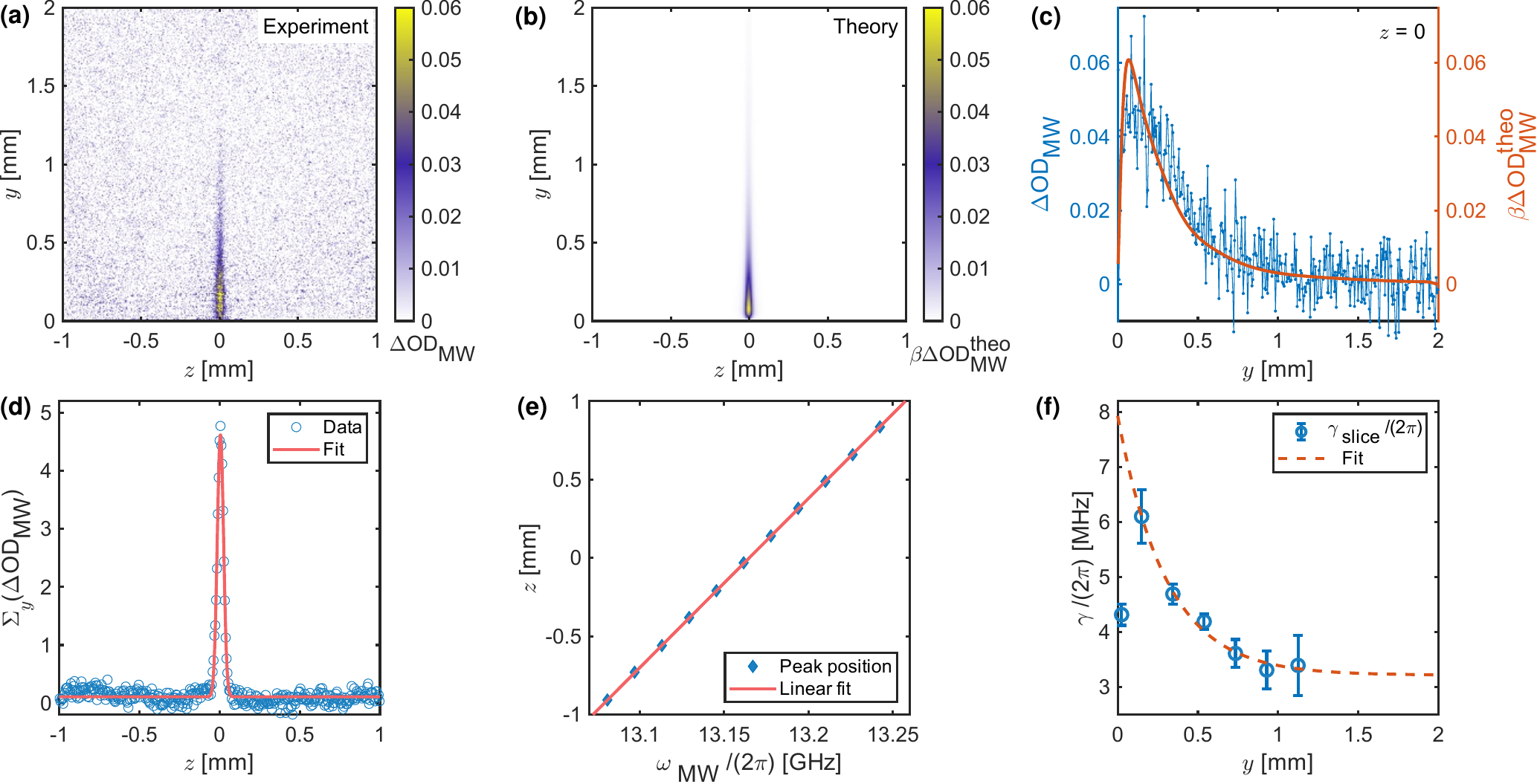}
\caption{\label{fig:single} (a) Atomic spectrum analyzer image of a single-tone MW signal at $\omega_\mathrm{MW}/(2\pi) = 13.165~\mathrm{GHz}$ and $P_\mathrm{MW}=0~\mathrm{dBm}$. The image shows $\Delta\textrm{OD}_\textrm{MW}$ recorded in the 2~mm cell at $T_\mathrm{set}=130.9(1)^\circ$C with $\tau_\textrm{MW} = 1~\mu s$ and averaged over $k\!=\!200$ repetitions in 2.2~s total integration time. The image ($369\times369$ pixels) is rotated by $\sim\! 1^{\circ}$ counterclockwise to align the signal with the image columns. (b) Theoretical simulation of the same image re-scaled by a global factor $\beta = 0.42$ to match the signal amplitude to the data. (c) Experimental $\Delta \textrm{OD}_\textrm{MW}$ at $z=0$ as a function of $y$ in comparison with the simulation. (d) Integrated signal $\Sigma_y(\Delta\textrm{OD}_\textrm{MW})$ with $y\in\left[ 0 , 1.2~\textrm{mm}\right]$. The Gaussian fit has a FWHM of $\Delta z_\textrm{FWHM} = 46(1)~\mu\textrm{m}$ corresponding to $\gamma/(2\pi) = 4.3(1)~\textrm{MHz}$.  (e) Signal peak position as a function of the applied MW frequency $\omega_\mathrm{MW}$ and linear fit. (f)
FWHM of Gaussian fits as in (d) to $195~\mu$m wide slices of data as a function of $y$ and exponential fit excluding the data point at $y=0$. The width plateaus at $\gamma/(2\pi) = 3.2(2)~\textrm{MHz}$. Error bars are one standard error (SE) confidence intervals of the fit parameters.}
\end{figure*} 

We used our atomic spectrum analyzer to record frequency spectra of a variety of MW test signals, which we present and analyze in the following. 
We start with measurements of a single-tone MW signal to characterize spectral resolution, bandwidth, linearity and sensitivity of the device. We then show spectra of multi-tone signals and a spectrogram of a frequency-swept signal.

\subsubsection{Lineshape and spectral resolution}

To characterize lineshape and spectral resolution, a single-tone microwave signal of angular frequency $\omega_\mathrm{MW}$ and power $P_\mathrm{MW}$ is applied and the corresponding $\Delta\textrm{OD}_\textrm{MW}$ image is recorded with the 2 mm cell, see Fig.~\ref{fig:single}(a). The interior bottom edge of the cell is at $y=0$ and the signal has the form of a thin line perpendicular to the chip surface, following a line of constant magnetic field strength and decaying with increasing distance from the CPW. The position of this line along $z$ yields the frequency of the MW signal and its width gives the spectral resolution.
Fig.~\ref{fig:single}(b) shows a theoretical simulation of the same image (see Sec.~\ref{sec5:simulations} below for details) for the same experimental conditions as in (a). It shows good agreement with the experimental signal if its amplitude is rescaled by a global factor $\beta=0.42$, which is likely explained by an uncertainty in the MW signal strength on the CPW at the vapor cell position due to standing waves and other propagation effects. Noise is not included in the simulation.

In Fig.~\ref{fig:single}(c) we show $\Delta\textrm{OD}_\textrm{MW}$ in the image column at $z=0$ as a function of $y$. The decay of $\Delta\textrm{OD}_\textrm{MW}$ for $y > 0.2~\mathrm{mm}$ reflects the decay pattern of the MW field distribution above the CPW, whereas the steep decrease close to the cell wall at $y<0.2~\mathrm{mm}$ is due to spin relaxation caused by wall collisions \cite{horsley2013imaging}. 
In the simulation described in Sec.~\ref{sec5:simulations}, both effects are accounted for, resulting in a quantitative match with the experimental data when the global rescaling factor $\beta$ is applied. 

In principle, one row of the $\Delta\textrm{OD}_\textrm{MW}$ image along $z$ can already provide the frequency spectrum of the MW signal. Since the image contains many such rows, it simultaneously yields many such spectra with different sensitivities depending on the distance from the CPW. In order to increase the signal-to-noise ratio, we sum over the image columns to obtain an integrated signal $\Sigma_y(\Delta\mathrm{OD}_\mathrm{MW})$ with $y\in\left[ 0 , 1.2\,\textrm{mm}\right]$. The resulting frequency spectrum of the single-tone signal is shown in Fig.~\ref{fig:single}(d). A Gaussian fit yields the position $z_{\omega_\mathrm{MW}}$ and the FWHM $\Delta z_\textrm{FWHM}$ of the peak, corresponding to $\omega_\mathrm{MW}$ and the frequency resolution $\gamma$, respectively.
The frequency axis of the spectrum is calibrated by scanning $\omega_\mathrm{MW}$ and recording $z_{\omega_\mathrm{MW}}$. The result is displayed in Fig.~\ref{fig:single}(e), along with the linear fit $\omega_\mathrm{MW}(z)/(2\pi) = 0.0926\,[\textrm{GHz}/\textrm{mm}]\times z + 13.165\,\textrm{GHz}$, which also allows us to determine the magnetic field gradient.

\begin{figure*}[ht]
\includegraphics[width=0.9\textwidth]{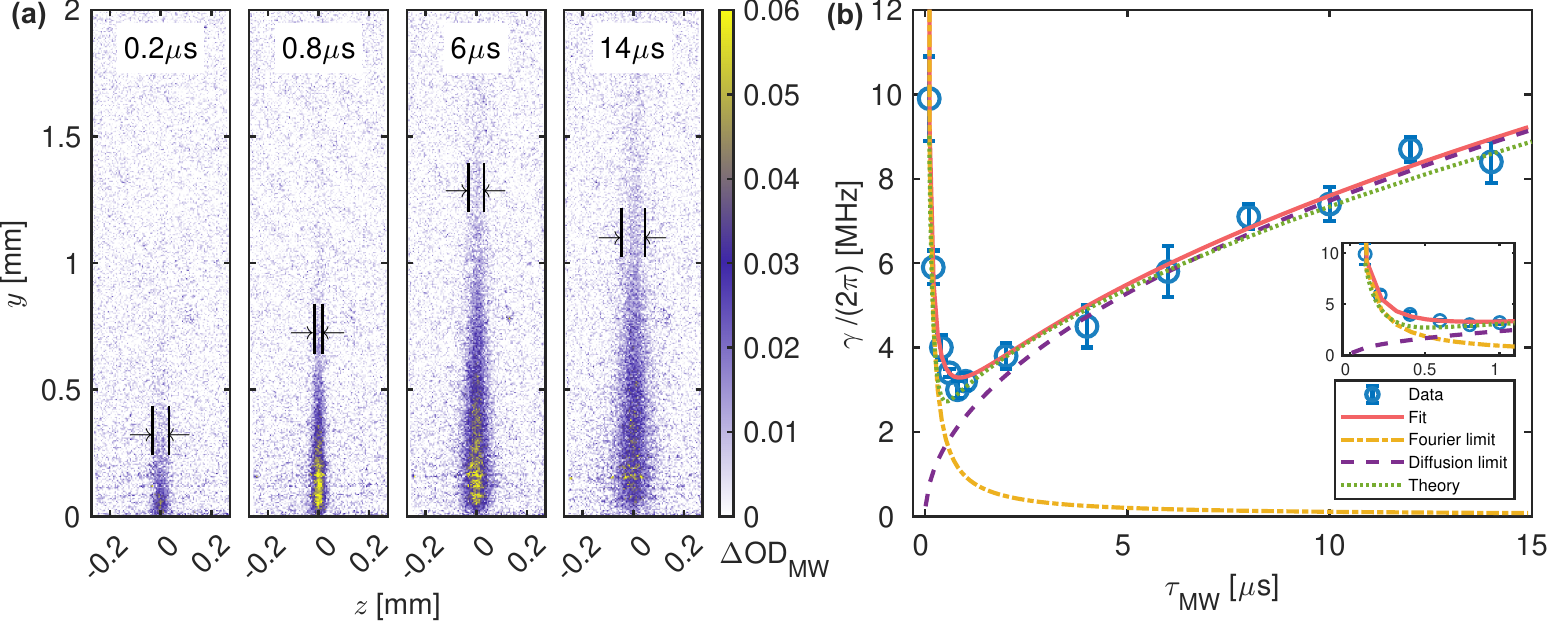}
\caption{\label{fig:FWHMvsMWTime} (a) Images of $\Delta\textrm{OD}_\textrm{MW}$ for different $\tau_\mathrm{MW}$ and other conditions as in Fig.~\ref{fig:single}(a). The signal width $\gamma$ was measured at the indicated positions where power broadening is negligible. (b) Signal width $\gamma$ as a function of $\tau_\mathrm{MW}$ (error bars: 1 SE confidence interval). Red solid line: fit (see text); yellow dash-dotted line: Fourier limit and intrinsic linewidth $\gamma_0$; purple dashed line: diffusion limit; green dotted line: simulation result. Inset: zoom into the region $\tau_\mathrm{MW}<1~\mu$s.}
\end{figure*}

In Fig.~\ref{fig:single}(f) we investigate the dependence of the signal width $\gamma$ on the distance from the CPW. To this end, the $\Delta\textrm{OD}_\textrm{MW}$ image is divided into horizontal slices that are 36 rows ($195~\mu$m) wide and integrated over. A Gaussian fit as in Fig.~\ref{fig:single}(d) yields $\gamma$ for each data slice, which is plotted in Fig.~\ref{fig:single}(f). Since the signal near $y=0$ is very small due to wall collisions, this data point is excluded. An exponential fit to the remaining data shows that $\gamma$ levels off at $\gamma/(2\pi)=3.2(2)~\mathrm{MHz}$ sufficiently far away from the CPW so that power broadening of the signal is negligible. We identify this value with the frequency resolution of our atomic spectrum analyzer.

In Fig.~\ref{fig:FWHMvsMWTime}(a) we study the dependence of the signal shape on MW pulse duration $\tau_\mathrm{MW}$. The length of the signal on the image increases with $\tau_\mathrm{MW}$ as the weaker MW field further away from the CPW also begins to change the atomic spin state. At the same time, the signal width increases due to atomic diffusion. The signal width $\gamma$ (plateau values without power broadening, analyzed as in Fig.~\ref{fig:single}(f)), is shown in Fig.~\ref{fig:FWHMvsMWTime}(b) as a function of $\tau_\mathrm{MW}$. The data is fitted with the heuristic model 
\begin{equation}
\label{eqn:tmwfit}
\gamma(\tau_\mathrm{MW}) = \frac{2\pi a}{\tau_\mathrm{MW}}+ A\sqrt{\tau_\mathrm{MW}} + \gamma_0,
\end{equation}
where $\gamma_0/(2\pi) = 13.7~\mathrm{kHz} $ is the intrinsic linewidth estimated from all relevant collisional relaxation processes \cite{horsley2015high}, and $a$ and $A$ are fit parameters. The first term of Eq.~(\ref{eqn:tmwfit}) accounts for the Fourier limit, which dominates the total width when $\tau_\mathrm{MW}$ is small. The fit yields $a=0.93(4)$, consistent with the Fourier width of the pulsed MW signal. 
The second term of Eq.~(\ref{eqn:tmwfit}) accounts for atomic diffusion and the fit yields $A/(2\pi) = 2.37($4$)\times 10^{9}  ~\mathrm{Hz}/\sqrt{\mathrm{s}}$. A simple model of atomic diffusion for a time $\tau_\mathrm{MW}$ allows us to calculate $A/(2\pi) = 2\sqrt{2\mathrm{ln}(2)}g\mu_B G\sqrt{2D}/h = 5.33\times 10^{9} ~\mathrm{Hz}/\sqrt{\mathrm{s}}$, with the diffusion coefficient of $^{87}$Rb in N$_2$ buffer gas being $D=2.53~\textrm{cm}^2/$s in the current cell at the experimental temperature. The experimentally obtained $A$ is smaller by a factor $2.2$, reflecting the fact that only atoms transferred at the beginning of the MW pulse diffuse for the entire $\tau_\mathrm{MW}$, while atoms transferred at the end of the pulse have no time to diffuse, resulting in a reduced average diffusion distance. 
Figure~\ref{fig:FWHMvsMWTime}(b) also shows $\gamma$ as obtained from the simulation in Sec.~\ref{sec5:simulations}, which agrees well with the data. In conclusion, there is an optimal MW pulse duration for a given magnetic field gradient in order to achieve the finest frequency resolution, which in this case is $3.0(1)$~MHz when $\tau_\mathrm{MW} = 0.8~\mu$s.

\subsubsection{Instantaneous bandwidth}

A key performance metric of our atomic MW spectrum analyzer is the instantaneous bandwidth $\Delta \omega_\mathrm{BW}$, which is primarily determined by the magnetic field gradient and cell length, see Eq.~(\ref{eq:bandwidth}). We measure $\Delta \omega_\mathrm{BW}$ by applying two microwave tones corresponding to the left and right edge of the cell, respectively. The results for our three MEMS atomic vapor cells of different length are presented in Table \ref{Tab}.
\begin{figure}[ht]
\includegraphics[width=0.48\textwidth]{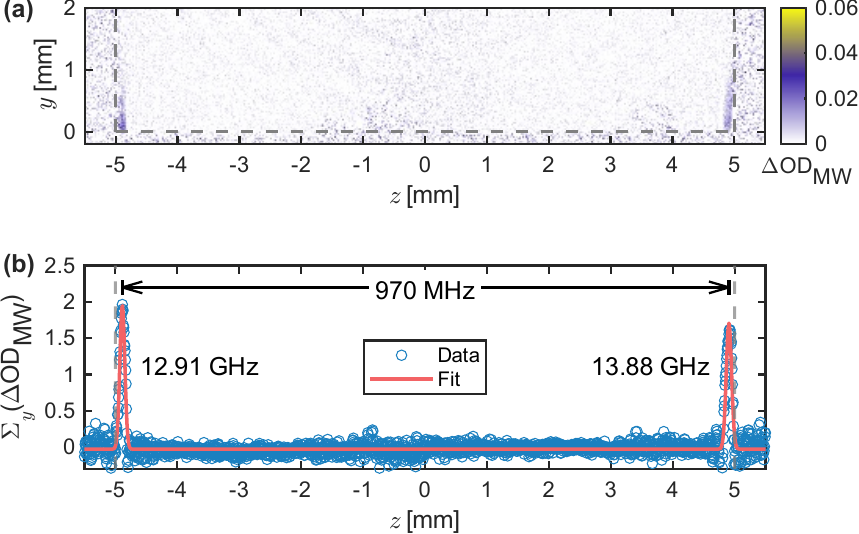}
\caption{\label{fig:10mmCell}
(a) Measurement of $\Delta \omega_\mathrm{BW}$ with the $L=10$~mm cell. The gray dashed lines indicate the cell edges. Here, $\tau_\mathrm{MW}= 2~\mu\mathrm{s}$ and the two microwave tones corresponding to the left and right cell edge are $\omega_\mathrm{MW}/(2\pi)=12.91~\mathrm{GHz}$ and $\omega_\mathrm{MW}/(2\pi)=13.88~\mathrm{GHz}$, respectively. The two tones are applied consecutively and the images are stitched together at $z=0$. (b) MW spectrum obtained by integrating the image in (a) over $y\in\left[ 0 , 1.2~\textrm{mm}\right]$, with Gaussian fits to the two MW signals close to the cell edges.}
\end{figure}
\begin{table}[ht!]
\caption{Instantaneous bandwidth and spectral resolution measured with three MEMS cells of different sizes and buffer gas pressures.}{\label{Tab}}
\begin{tabular}{ c | c | c | c  }
\hline
 $L$ [mm] & $P_\mathrm{N_2}^\mathrm{fill}$ [mbar] & $\gamma/(2\pi)$ [MHz] & $\Delta \omega_\mathrm{BW}/(2\pi)$ [MHz]\\ 
\hline
 2 & 127 & 3.0(1) & 184 \\ 
 6 & 112 & 5.7(2) & 550\\  
 10 & 100 & 9.3(6) & 970 \\
 \hline
\end{tabular}
\end{table}
\begin{figure*}[ht]
\includegraphics[width=0.9\textwidth]{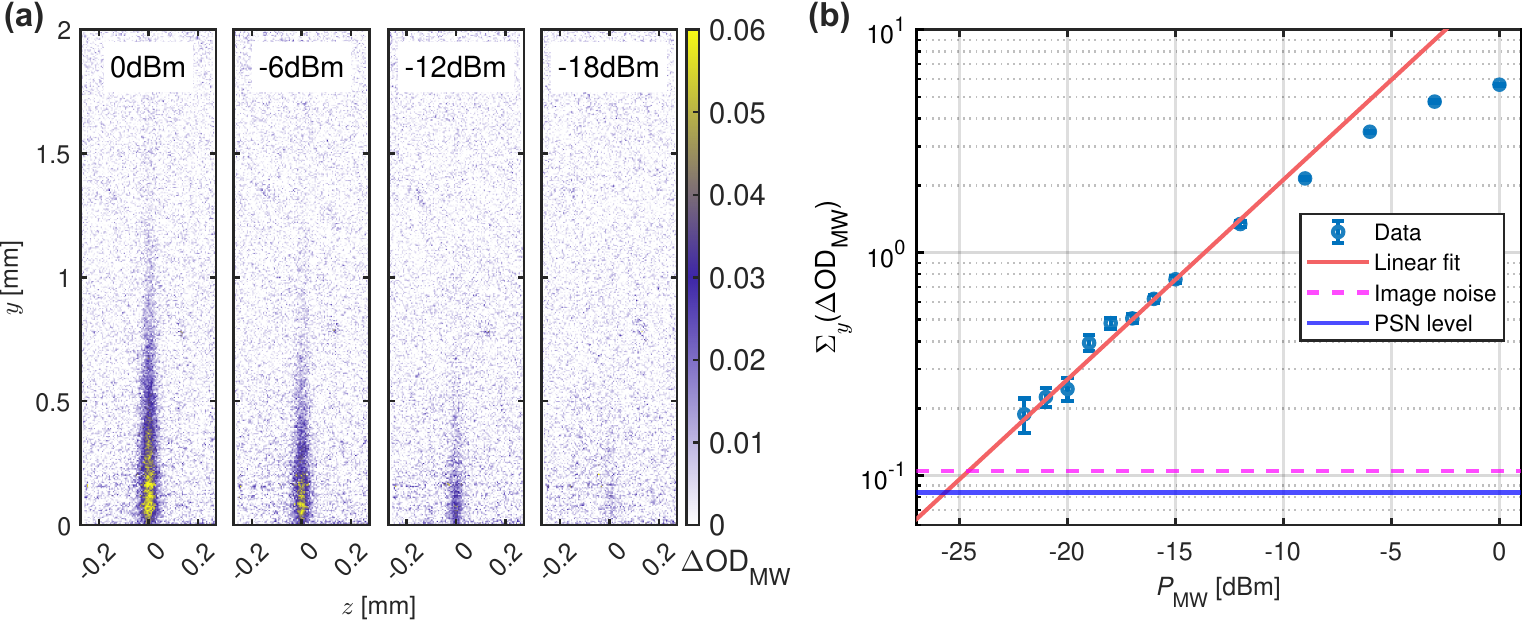}
\caption{\label{fig:MWPower} 
(a) Images of $\Delta\textrm{OD}_\textrm{MW}$ for single-tone MW signals of different power $P_\mathrm{MW}$. Here, $\tau_\mathrm{MW}= 2~\mu\mathrm{s}$ and other conditions are as in Fig.~\ref{fig:single}(a). (b) Peak values of the Gaussian fit to the integrated spectrum $\Sigma_y(\Delta\mathrm{OD}_\mathrm{MW})$ with $y\in\left[ 0 , 1.2\,\textrm{mm}\right]$, as a function of $P_\mathrm{MW}$ (error bars: 1 SE confidence interval). The linear fit is performed on the leftmost 9 points, the magenta dashed line is the noise level without MW, and the blue solid line is the photon shot noise level.}
\end{figure*}
The exact positions of the three cells with respect to the permanent magnets are not identical, resulting in slightly different magnetic field gradients. 
The longest MEMS cell with $L=10$~mm reaches a maximal instantaneous bandwidth of $\Delta \omega_\mathrm{BW}/(2\pi)\approx 1.0$~GHz, see Fig.~\ref{fig:10mmCell}.
Note that the frequency resolution values for the $L=6$~mm and $L=10$~mm cells may not be optimal since the MW pulse duration was not adjusted. 
To further increase the instantaneous bandwidth, both the magnetic field gradient and the cell length can be increased.

\subsubsection{Linearity and dynamic range}

We now explore the response of the atomic spectrum analyzer to MW signals of different power $P_\mathrm{MW}$ generated by the signal generator, see Fig.~\ref{fig:MWPower}(a). As before, we sum over the image column values to obtain $\Sigma_y(\Delta\mathrm{OD}_\mathrm{MW})$ as a function of $z$ and then fit it with a Gaussian to obtain the peak value of the spectrum. This peak value shows a linear dependence on $P_\mathrm{MW}$ for small powers, see Fig.~\ref{fig:MWPower}(b). At higher MW powers it begins to saturate, since the atomic response in the lower part of the $\Delta\textrm{OD}_\textrm{MW}$ image, corresponding to the region closer to the CPW, shows saturation and power broadening, see the two leftmost images in Fig.~\ref{fig:MWPower}(a). The nonlinearity of the MW pre-amplifier used throughout the experiment also contributes to the saturation.

An advantageous feature of our spectrum analyzer based on an imaging scheme is the increased dynamic range due to the spatially inhomogeneous MW field of the CPW, which enables the simultaneous detection of weak and strong MW signals in different parts of the spectrum. Weak signals are most sensitively detected close to the CPW, while at the same time strong signals at other frequencies can be detected with similarly high spectral resolution in the imaging region far from the CPW.

\subsubsection{Sensitivity}

The minimum MW power in the measurement of Fig.~\ref{fig:MWPower}(b) is $P_\mathrm{MW}=-22$~dBm, detected in $2.2$~s integration time with signal-to-noise ratio $\mathrm{SNR}=1.7$. Here, the noise is obtained from a similar image without applied MW signal. Normalized to $1$~s integration time and $\mathrm{SNR}=1$ this corresponds to a detection limit of $P_\mathrm{MW}=-23$~dBm, where $P_\mathrm{MW}$ refers to the input coaxial cable of the current setup. We also estimate the expected noise on $\Sigma_y(\Delta\mathrm{OD}_\mathrm{MW})$ due to photon shot noise (PSN), which corresponds to a minimum detectable MW power of $\sim -24$~dBm in $1$~s integration time under the current experimental conditions, see Fig.~\ref{fig:MWPower}(b). The PSN on the images could be decreased by employing higher probe power, since the camera full well capacity is not exhausted by the current setting. However, this would also cause re-pumping of the vapor by the probe light, decreasing the $\Delta\mathrm{OD}_\mathrm{MW}$ signal. The probe power in the experiment is a compromise between these effects. MW power sensitivity is also limited by the low duty cycle of the detection system, with camera digitization consuming most of the frame time ($\sim\!4$ ms for the current image size). Higher sensitivities are envisaged for improved experimental parameters, see Sec.~\ref{sec6:discussion} below.

\begin{figure*}[htp]
\includegraphics[width=\textwidth]{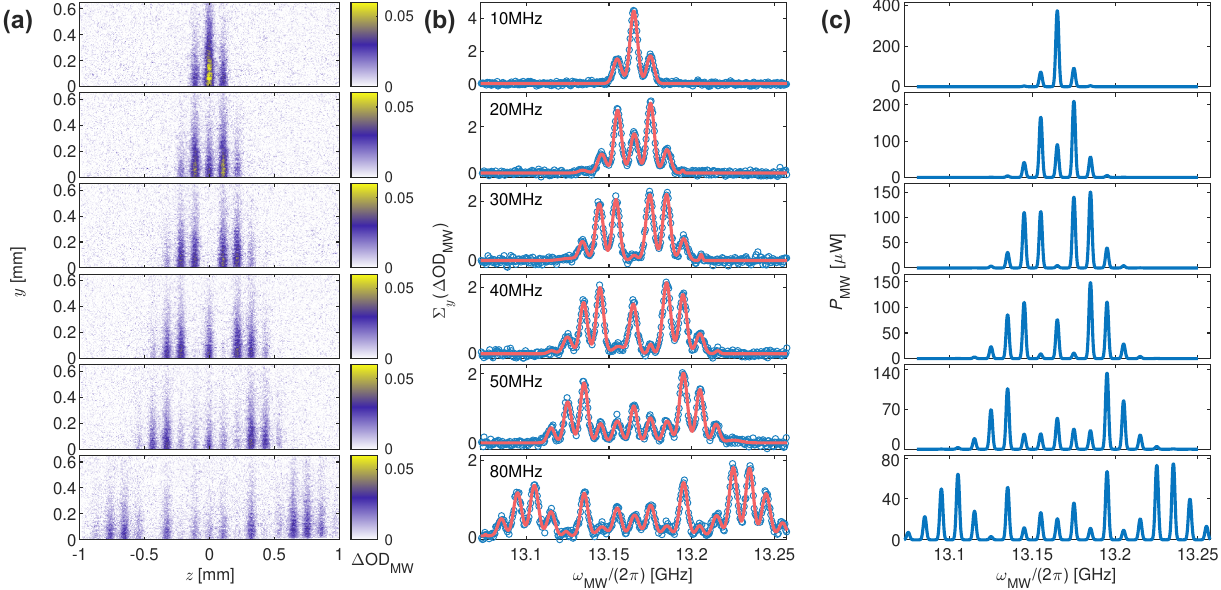}
\caption{\label{fig:FM} Atomic spectrum analyzer images of different frequency-modulated MW signals with carrier frequency $13.165~\mathrm{GHz}$, modulation frequency $\omega_\mathrm{mod}/(2\pi) = 10~\mathrm{MHz}$, and modulation index increasing from top to bottom. (a) Images of $\Delta\textrm{OD}_\textrm{MW}$ recorded with $\tau_\mathrm{MW}=3~\mu$s. (b) Corresponding integrated spectra $\Sigma_y(\Delta\mathrm{OD}_\mathrm{MW})$ with multi-peak Gaussian fits. The frequency deviation $\omega_\Delta/(2\pi)$ of each spectrum is indicated. (c) Frequency spectra of the same signals after the CPW recorded with a commercial electronic spectrum analyzer. }
\end{figure*}
\subsubsection{Frequency spectra of multi-tone signals}

To demonstrate the operation of the atomic spectrum analyzer with more complex signals, we apply different frequency-modulated MW signals, see Fig.~\ref{fig:FM}. The carrier signal is set to $\omega_c/(2\pi) = 13.165$~GHz and its frequency is modulated with a $\omega_\mathrm{mod}/(2\pi)=10$~MHz sinusoidal baseband signal. 
Increasing the frequency deviation $\omega_\Delta$ increases the modulation index, resulting in more sidebands with varying amplitudes. The situation in which the carrier signal is surpassed ($\omega_\Delta > \omega_\mathrm{mod}$) is of interest in communications, since the energy is not consumed by the idle carrier. The measurements were performed with $\tau_\textrm{MW} = 3~\mu\mathrm{s}$ to reduce the Fourier linewidth. The obtained images and MW frequency spectra agree quantitatively with corresponding measurements made by a commercial electronic spectrum analyzer.

\subsubsection{Spectrogram of a frequency-swept signal}

\begin{figure}[htp]
\includegraphics[width=0.5\textwidth]{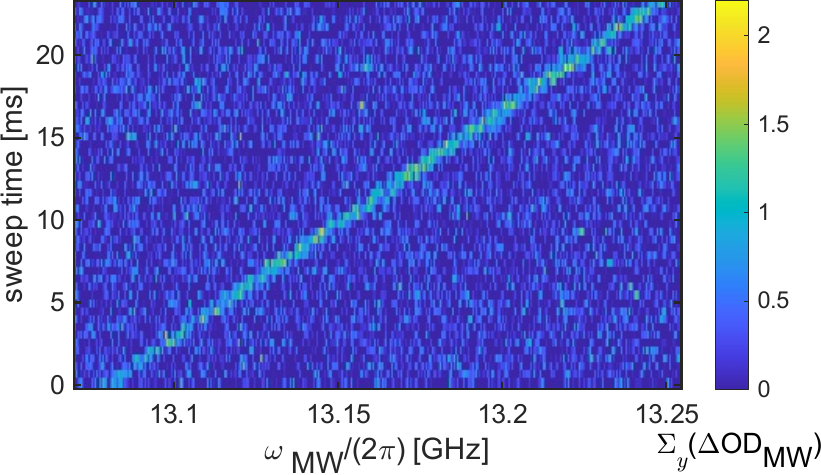}
\caption{\label{fig:spectrogram} Spectrogram of frequency-swept MW signal. Each line represents the integrated signal $\Sigma_y(\Delta\mathrm{OD}_\mathrm{MW})$ over the selected 20-row-image measured within $480~\mu$s. With 50 measurements during the $24$~ms sweep time, all timings are confirmed by a separate oscilloscope triggered by the camera exposure pulse. The measurement confirms the linearity of the MW frequency sweeping process.}
\end{figure}

To demonstrate the temporal resolution of the atomic spectrum analyzer, we record a spectrogram of a frequency-swept MW signal, see Fig.~\ref{fig:spectrogram}. The signal, generated by a synthesized MW sweeper, sweeps linearly from $13.075$~GHz to $13.25$~GHz over $24$~ms with an output power of $10$~dBm. We use the same sequence as in Fig.~\ref{fig:principle}(d) with $2~\mu$s probe pulse duration, $\tau_\mathrm{MW}=2~\mu$s, and $k=1$. The time resolution is currently limited by the camera frame rate, which is proportional to the number of selected image rows. By reading out only 20 rows of the $\Delta\mathrm{OD}_\mathrm{MW}$ image (close to the bottom edge of the cell), we achieve a frame time of $240~\mu$s, yielding a time resolution of $480~\mu$s, equivalent to a frame rate of approximately $2$~kHz.
Future experiments with a faster camera (or a lock-in camera) could achieve a time resolution of less than $10~\mu$s by optimizing pulse parameters and experimental settings.  

\section{\label{sec5:simulations}Theoretical simulations}
We simulate the atomic spectrum analyzer images in order to understand the physical processes in greater detail and to quantitatively compare with the data. Our numerical simulation models the dynamics of a $^{87}$Rb atomic vapor in a two-dimensional, thin cell, taking into account the Hamiltonian of all 8 levels of the $5^2\mathrm{S}_{1/2}$ ground state and all 16 levels of the $5^2\mathrm{P}_{3/2}$ excited state \cite{steck_rubidium_2021}. We consecutively model the steps of our measurement sequence, consisting of optical pumping, microwave interaction, and laser absorption imaging, taking into account atomic diffusion and collisional relaxation.

\textit{Optical pumping.} A thorough description of optical pumping in atomic vapors with diffusing atoms is given in \cite{appelt_theory_1998}. There, the equations of motion are solved in Liouville space, while here we rely on a rate equation approach \cite{atoneche_simplified_2017}. By assuming that the optical coherences of the density matrix of the $^{87}$Rb atoms quickly reach a steady state, and the decay rates from the excited states are much higher than the excitation rates, we can describe the optical pumping process with partial differential equations for the populations $p_n \equiv p_n(t, z, y)$, where $n=1,...,8$:
\begin{equation}
\frac{\partial}{\partial t} p_n = \sum_{j = 1}^{8} a_{nj} p_j + \left(\sum_{j = 1}^{8} \frac{\gamma_0}{8} p_j - \gamma_0 p_n\right) + D \nabla^2 p_n.
\label{eqn:rate_eq}
\end{equation}
The coefficients $a_{nj}$ are given by the excitation rates from a ground state $j$ to the excited states (proportional to the pump beam intensity), multiplied by the decay rate of each excited state to a particular ground state $n$ \cite{atoneche_simplified_2017}. The excitation rates are determined by the solution of the optical Bloch equations for each pair of ground state and excited state in the steady state approximation. In our MEMS atomic vapor cell, the decay rate is mainly determined by the quenching process with the $\mathrm{N}_2$ buffer gas (which approximately preserves the nuclear spin). The second term in Eq.~(\ref{eqn:rate_eq}) describes ground-state relaxation due to collisions, where wall collisions along the $x$-direction dominate, and the third term describes atomic diffusion. It is possible to solve these equations by expanding $p_n$ into a series of diffusion modes, because the coefficients $a_{nj}$ are approximately position-independent in our experiment. This expansion turns the partial differential equations into rate equations for each diffusion mode, which we solve numerically using a matrix exponential ansatz. For the initial conditions, we assume equal populations in each ground state prior to pumping.

\textit{Microwave interaction.} For modelling the microwave interaction, we assume that the static magnetic field is sufficiently strong so that the microwave only resonantly interacts with one pair of ground states, see Fig.~\ref{fig:principle}(c). In this case, we can simulate the process by solving the Bloch-Torrey equations \cite{torrey1956bloch,kosugi_theory_2005} based on the two sub-levels $p_i$, $p_j$ that are involved in the MW interaction with added diffusion and relaxation terms:
\begin{align}
\dot{u} &= - \delta v - \gamma_0 u + D \nabla^2 u,\\ 
\dot{v} &= \delta u - \Omega w - \gamma_0 v + D \nabla^2 v,\\ 
\dot{w} &= \Omega v - \gamma_0 w + D \nabla^2 w,\\ 
\dot{S} &= - \gamma_0 S + \gamma_0 / 4 + D \nabla^2 S.
\end{align}
The parameters $u$ and $v$ are defined using the microwave coherences in a typical manner \cite{kosugi_theory_2005}, and $w$ and $S$ are related to the populations of the these two states via $p_{i/j} = \frac{1}{2} (S \pm w)$. 
The position-dependent Rabi frequency $\Omega = \Omega(y)$ is calculated based on a simulation of the microwave magnetic field distribution with COMSOL (blue fan-shaped contour in Fig.~\ref{fig:principle}(a)) using the input MW power from the experiments. The position-dependent detuning $\delta = \delta(z)$ is given by the magnetic field gradient. We solve the Bloch equations numerically \cite{py-pde} with the boundary conditions $p_n(t)|_\mathrm{walls} = 1/8$ in the rectangular cell.

\begin{figure*}[htp]
\includegraphics[width=0.6\textwidth]{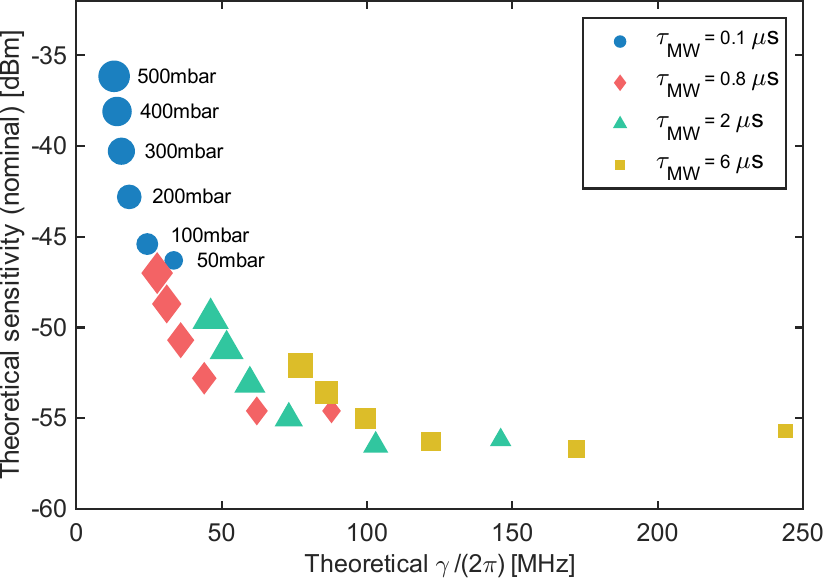}
\caption{\label{fig:prediction} Theoretical dependence of the FWHM frequency resolution and the nominal MW power sensitivity on the MW pulse length and buffer gas pressure. The nominal MW power sensitivity is estimated for the current MW setup and normalized to $1$~s integration time.}
\end{figure*}

\textit{Laser absorption.} We calculate the absorption of the probe pulse using the same procedure as described in \cite{zentile_elecsus_2015}. We also approximate the pumping effect of the probe beam on the vapor, by calculating the time-dependent populations using similar rate equations as Eq.~(\ref{eqn:rate_eq}), setting the initial conditions to be the populations after the optical pumping and the microwave interaction. 

By plugging in all the experimental parameters, we calculate the evolution of populations for all states after each of the optical pumping, MW interaction and probe absorption processes, for each pixel on the grid. With the Rb number density at the experimental temperature and the cell length, we can then calculate the evolution of the OD images after each process. By comparing the OD image obtained from the same procedures but with MW turned off, we can finally generate an image of $\Delta\textrm{OD}_\textrm{MW}^\textrm{theo}$ as shown in Fig.~\ref{fig:single}(b) and (c). Similarly, by varying the MW pulse duration $\tau_\mathrm{MW}$ in the simulation, we derive the dependence of the spectral resolution on $\tau_\mathrm{MW}$, shown as dotted line in Fig.~\ref{fig:FWHMvsMWTime}(b).

The theoretical simulation adequately reproduces the signal spatial shape and spectral resolution, which are dominated by the well-known diffusion process and the precisely calibrated static magnetic field gradient along $z$. However, a notable discrepancy persists between the measured and simulated absolute values of the $\Delta\textrm{OD}_\textrm{MW}$ image, which differ by the global factor $\beta$ introduce in Sec.~\ref{sec4:results} (compare Figs.~\ref{fig:single}(a) and (b)), suggesting a systematic effect uniformly affecting all pixels. 
We suspect that this discrepancy arises from the inaccurate estimation of the MW field strength at the vapor cell position along the CPW, due to propagation effects and possible standing waves along the CPW signal wire.

\section{\label{sec6:discussion}future improvements}

We discuss possible improvements to our atomic spectrum analyzer based on theoretical simulations with the model described in the previous section. 
To increase the instantaneous bandwidth while maintaining a practical cell length, a larger static field gradient is required. Considering a cell with $L=2$~cm and $t_1=675~\mu$m thickness in a realistic magnetic field gradient of $G=50~\mathrm{T}/\mathrm{m}$ \cite{marteau2017variable}, the instantaneous frequency span exceeds $25$~GHz. Using this gradient and cell dimension in our theoretical simulation, we calculate the dependence of the frequency resolution $\gamma/(2\pi)$ and MW power sensitivity (normalized to $1$~s integration time) on MW pulse length $\tau_\mathrm{MW}$ and buffer gas pressure $P_\mathrm{N_2}$, see Fig.~\ref{fig:prediction}. For each set of parameters, we employ an optimized algorithm to enhance SNR for weak MW signals by selecting an effective image region. MW power sensitivity is defined as the equivalent photon-shot noise level over the selected pixels at optimal probe power.

Fig.~\ref{fig:prediction} shows that higher buffer gas pressures improve frequency resolution by decreasing diffusion distance, but at the same time decrease sensitivity, because larger collisional broadening decreases the initial optical depth $\mathrm{OD}$ as well as the value after optical pumping $\Delta\mathrm{OD}_\mathrm{pump}$. Longer MW pulses improve sensitivity but also reduce frequency resolution (compare Fig.~\ref{fig:FWHMvsMWTime}(b)). For $\tau_\mathrm{MW}=0.8~\mu$s and $150$~mbar $\mathrm{N}_2$ buffer gas, a frequency resolution of $50$~MHz is possible, meaning that $\approx 500$ frequency channels can be simultaneously detected over the $25$~GHz span, with single-channel power sensitivity of $-55$~dBm. 

Higher probe beam powers would be desirable to further enhance sensitivity by reducing photon shot noise, however, the probe power is constrained by the optical pumping due to the resonant probe beam. Alternative detection schemes, such as a Faraday rotation imaging scheme \cite{gajdacz2013non,kiehl2023coherence} with an off-resonant probe beam, are interesting to explore in this context. Probing of atomic vapors by Faraday rotation is routinely employed in state-of-the-art atomic magnetometers for DC or low-frequency RF fields \cite{kominis2003subfemtotesla,savukov2005tunable} and can reach atomic projection noise limited performance \cite{wasilewski2010}. Cameras with higher frame rates and larger full well capacity or an analog photodetector array with many channels would enable operation at larger probe beam powers, thus improving MW sensitivity significantly. Moreover, Faraday detection schemes are compatible with higher buffer gas pressures and higher vapor temperature \cite{kominis2003subfemtotesla}, allowing for improved frequency resolution without compromising sensitivity.

At very large static field gradients, uniquely identifying the frequencies in a multi-tone MW signal is complicated by the fact that a single MW frequency can drive multiple MW transitions between different ground state pairs at different positions in the cell, resulting in multiple lines on the atomic spectrum analyzer image for each frequency component. One approach to avoid this complication is to fully polarize atoms into a single magnetic sub-level, suppressing other MW transitions and enabling clear discrimination across a broad frequency range. Alternatively, pre-calibrating these signal lines based on distinct transition strengths \cite{horsley_frequency-tunable_2016} for a given gradient allows for unique post-identification of the input MW frequency using an inverse method.

Real-time operation is a crucial capability for spectrum analyzers, promising gap-less coverage of all incident broadband MW events in the time domain. Our theoretical model suggests that an efficiently optically pumped steady state can be established within microseconds with higher pump power. Therefore, operating multiple atomic spectrum analyzers simultaneously allows for the design of experimental sequences to consecutively allocate their MW detection windows in the time domain, achieving a high probability of interception for rapidly changing events. Exploiting the fact that the atoms do not significantly attenuate the analyzed MW signal, multiple atomic spectrum analyzers can be connected in series for real-time operation or even broader spectral coverage. This would turn the anticipated $25$~GHz instantaneous bandwidth into a corresponding real-time analysis bandwidth.

Although our atomic spectrum analyzer is still in a proof-of-principle development stage, it is interesting to compare the present and anticipated performance with state-of-the-art electronic spectrum analyzers. Commercial real-time spectrum analyzers currently offer up to 2~GHz real-time bandwidth. These analyzers rely on time-overlapped FFT configurations, requiring powerful digital processing engines and large memory capacities. Users can choose from several options for frequency resolution (i.e.\ resolution bandwidth) within a given real-time span. This selection impacts the minimum measurement duration (time resolution) and noise level. Finer frequency resolution requires longer measurement durations but achieves lower noise levels. Typical values for frequency resolution range from sub-MHz to about 100~MHz, with minimum measurement durations from several hundreds of microseconds to several microseconds. The noise level for each resolvable frequency channel correspondingly ranges from -110~dBm to -60~dBm. 
In our current demonstration, the atomic spectrum analyzer already reaches $1$~GHz instantaneous bandwidth, while the improvements outlined above will make it possible to reach $25$~GHz real-time bandwidth at a resolution of $50$~MHz.
The minimum measurement duration is a few microseconds.
Our experimental demonstration has a sensitivity of $-23$~dBm in $1$~s integration time, while for the improved design we expect a sensitivity of $-55$~dBm in $1$~s integration time, which could be enhanced further by improving the optical probing of the atoms. The main promise of our approach is thus in increasing the real-time bandwidth beyond the capabilities of electronic spectrum analyzers, while only minimally perturbing the signal so that it can be further used while being analyzed.

\section{\label{sec7:conclusion}Conclusion}

Using a MEMS atomic vapor cell in a strong magnetic field gradient, we present for the first time an atomic spectrum analyzer for MW signals. Our system achieves $\approx 1$~GHz instantaneous bandwidth, $3$~MHz frequency resolution, $-23$~dBm MW power sensitivity, and $2$~kHz refresh rate. We demonstrate its capabilities by measuring FM signal spectra and recording frequency-swept MW signal spectrograms. A theoretical simulation shows good agreement with the measurements. Based on the simulation, an instantaneous bandwidth exceeding $25$~GHz seems feasible in an improved setup with larger field gradient. Further investigation indicates that utilizing Faraday imaging, increasing cell temperature, and enhancing buffer gas density could realise a high sensitivity atomic spectrum analyzer with improved frequency resolution. Our compact setup, based on common table-top optics and electronic components, can be integrated into small physics packages using optical waveguide or photonic integrated circuit technology. Compared to conventional methods, our approach promises broader instantaneous bandwidth, facilitating applications such as UWB signal detection, frequency-hopping communication, and real-time MW event monitoring.

\section*{Acknowledgements}
The authors would like to acknowledge Katja Glas for her contributions in making this work possible, Thomas Overstolz (CSEM) for the MEMS atomic vapor cells fabrication and Miguel Bueno Diez for checking the CPW near-field simulations. The authors thank Christoph Affolderbach, Stefan Bauer, Nigel Lawton, Paul Renardy, Gian-Luca Schmid, Vishal Shah, Jens Steinwandt and Tilman Zibold for fruitful discussions. This project has received funding from the European Union's Horizon 2020 research and innovation programme under grant agreement No. 820393 (MACQSIMAL).

\bibliography{AtomicSpectrumAnalyzer_Main_FinalSubmission}

\end{document}